\documentclass[preprint,12pt,3p]{elsarticle}

\usepackage{amssymb}
\usepackage{hyperref}
\usepackage{multirow}

\usepackage{gensymb}
\usepackage{graphicx}
\usepackage{caption}
\usepackage{subcaption}
\usepackage{amsmath}
\usepackage[capbesideposition={right,top}, facing=yes,capbesidesep=quad]{floatrow}
\journal{Medical Image Analysis}

\begin{document}
\begin{frontmatter}

\title{\textbf{Data Augmentation for Skin Lesion using Self-Attention based Progressive Generative Adversarial Network}}

\author[label1]{Ibrahim Saad Ali\corref{cor1}}
\address[label1]{Department of Computer Science, Faculty of Computer and Information, Assiut University, Assiut, 71511}

\cortext[cor1]{I am corresponding author}

\ead{ibrahim.cs@aun.edu.eg}

\author[label1]{Mamdouh Farouk Mohamed}
\ead{mamfarouk@aun.edu.eg}

\author[label1]{Yousef Bassyouni Mahdy}
\ead{mahdy@aun.edu.eg}

\begin{abstract}
	
Deep Neural Networks (DNNs) show a significant impact on medical imaging. One significant problem with adopting DNNs for skin cancer classification is that the class frequencies in the existing datasets are imbalanced. This problem hinders the training of robust and well-generalizing models. Data Augmentation addresses this by using existing data more effectively. However, standard data augmentation implementations are manually designed and produces only limited reasonably alternative data. Instead, Generative Adversarial Networks (GANs) is utilized to generate a much broader set of augmentations. This paper proposes a novel enhancement for the progressive generative adversarial networks (PGAN) using self-attention mechanism. Self-attention mechanism is used to directly model the long-range dependencies in the feature maps. Accordingly, self-attention complements PGAN to generate fine-grained samples that comprise clinically-meaningful information. Moreover, stabilization technique was applied to the enhanced generative model. To train the generative models, ISIC 2018 skin lesion challenge dataset was used to synthesize highly realistic skin lesion samples for boosting further the classification result. We achieve accuracy of 70.1\% which is 2.8\% better than the non-augmented one of 67.3\%.

\end{abstract}


\end{frontmatter}


\section{Introduction}
\label{introduction_sec}
Currently, Deep Neural Networks (DNNs) accomplish the state-of-the-art results across a variety of areas. Furthermore, DNNs have shown a significant impact on medical imaging by achieving a high accurate classification of many diseases, including skin cancer \cite{esteva2017dermatologist, fornaciali2016towards, perez2018data, valle2017data}. One significant problem with adopting DNNs for skin cancer classification is that there is a lack of labeled data leading to skewed class distributions. As a result, the class frequencies in the existing medical image datasets are imbalanced. This problem hinders the generalization of trained DNN models and resulting in biased DNN models towards the dominant classes in existing datasets. In the ISIC2018 challenge, there is a total of 10015 skin lesion images and the class distribution is heavily skewed among the seven types of skin lesions. Therefore, the state-of-the-art approaches for skin lesion classification and segmentation rely on heavy data augmentation \cite{yu2016automated, matsunaga2017image}.
\\ \indent
Data Augmentation alleviates the lack of labeled data by using existing data more effectively. It applies various transformations \cite{krizhevsky2012imagenet} to the original dataset to increase both the amount and diversity of data. These transformations include flips, rotations, random translations and addition of Gaussian noise. Data Augmentation is a vital technique, not only for imbalanced data but for any size of dataset e.g., the largest datasets such as Imagenet \cite{deng2009imagenet}, since data augmentation assists DNNs to exploit invariances in the existing data which leads to the training of robust and well-generalizing models. 
\\ \indent
Although data augmentation is an effective technique for improving the accuracy of DNNs, it has two potential issues. First, you need to search for improved data augmentation policies based on your understanding of the existing dataset. For example, labels of datasets such as handwritten characters dataset should be invariant to small shifts in location, small rotations or shears, changes in intensity, stroke thickness and size, etc. All these transformations lead the generated samples to be recognizable as a valid data sample when mixing these samples with the same label in feature space. Second, even when augmentation improvements have been found for a particular dataset, they may not be transferred to other datasets as effectively. For example, rotation of images during training is an effective data augmentation technique on CIFAR-10 \cite {krizhevsky2009learning}, but not on MNIST \cite {lecun1998mnist}, since the classifier will be unable to distinguish properly between handwritten “6” and “9” digits.

To bypass data augmentation issues, we propose using Generative Adversarial Networks (GANs) \cite{goodfellow2014generative} to automatically learn improved invariance space, in order to generate sample that preserves the class labels. The potential of GANs is hugged and scoped in their attempts to model the real image distribution by forcing the synthesized samples to be indistinguishable from real images. Based on these generative models, First successful attempts for medical data augmentation using GANs have been made in \cite{antoniou2017data, frid2018synthetic} at a level of small patches. Regarding skin cancer classification, skin images should be in a high resolution to spot malignancy markers that differ a benign from a malignant skin lesion. Very few works have shown outstanding results for high-resolution image generation. For example, progressive growing of GANs (PGAN) \cite{karras2017progressive} generates celebrity faces up to 1024 $\times$ 1024 pixels. The underlying idea is to start feeding the network with low-resolution samples and then progressively increases the resolution of generated images by gradually adding new layers to the generator and discriminator networks leading to increased stability in training behavior and very realistic, synthetic images at resolutions up to 1024 $\times$ 1024 pixels.

\begin{figure}[!ht]
	\centering
	\captionsetup{width=\columnwidth}
	\includegraphics[width=\columnwidth]{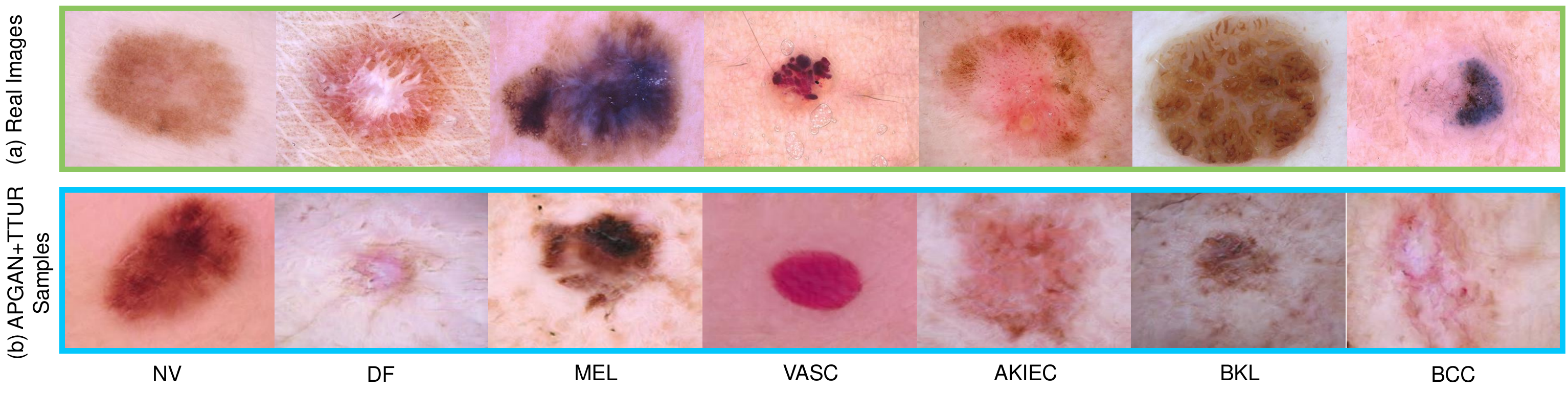}
	
	\caption{Real Images and APGAN+TTUR samples of ISIC 2018 dataset. Nevus(NV), Dermatofibroma(DF), Melanoma(MEL), Pigmented Bowen’s(AKIEC), Pigmented Benign keratoses(BKL), Basal cell carcinoma(BCC), and Vascular(VASC).  }\label{fig:APGAN_gen_imgs}
\end{figure}

In this paper, we propose a novel enhancement to PGAN using self-attention mechanism for generating high-definition, visually-appealing and clinically-meaningful synthetic skin lesion images. To the best of our knowledge, this work is the first that successfully incorporates the self-attention mechanism to PGAN for increasing the perceptual quality of the images by modelling the attention-driven long-range dependencies. Moreover, the Two Time Update Rule (TTUR) (imbalanced learning rate) is used to improve the network stability at high resolution $256\times 256$ pixels. The ISIC2018 challenge public dataset \cite{tschandl2018ham10000, codella2018skin} is used to train PGAN, attention progressive growing of GANs (APGAN) and APGAN with TTUR (APGAN+TTUR). The generated samples of APGAN+TTUR is illustrated in Figure \ref {fig:APGAN_gen_imgs}. To finely separate the best performance GAN, we use GAN-train and GAN-test of \cite { shmelkov2018good} and afterwards the best performance GAN was used to augment the training data of ISIC2018 dataset.  Experiments show that our method can improve the classification accuracy by 2.8\% on average.

Briefly the key contributions of this paper are:
\begin{enumerate}
	\item  Novel enhancement using self-attention based progressive Generative adversarial network.
	\item Apply a stabilized training procedure for increased stability of training behavior. 
	\item Generate high-definition, visually-appealing and clinically-meaningful skin lesion images.
	\item  Improve classification accuracy over the corresponding real-only and standard augmentation counterparts.
\end{enumerate}

The remainder of this paper is organized as follows: In Section \ref{sec:Proposed_Approach_sec} briefly recapitulates the APGAN framework as well as the stabilization technique. Section \ref{sec:Experiment} outlines our methodology that builds upon previously published literature and discuss the results of our experiments in detail. In addition, it discusses artifacts of the generated samples and the utilizing of APGAN+TTUR as an augmenter to boost the classification accuracy. Finally, in the Conclusion, we conclude our work and give an outlook on future work.

\section{Proposed Approach}
\label{sec:Proposed_Approach_sec}

\begin{figure}[!ht]
	\centering
	\captionsetup{width=\columnwidth}
	\includegraphics[width=\columnwidth]{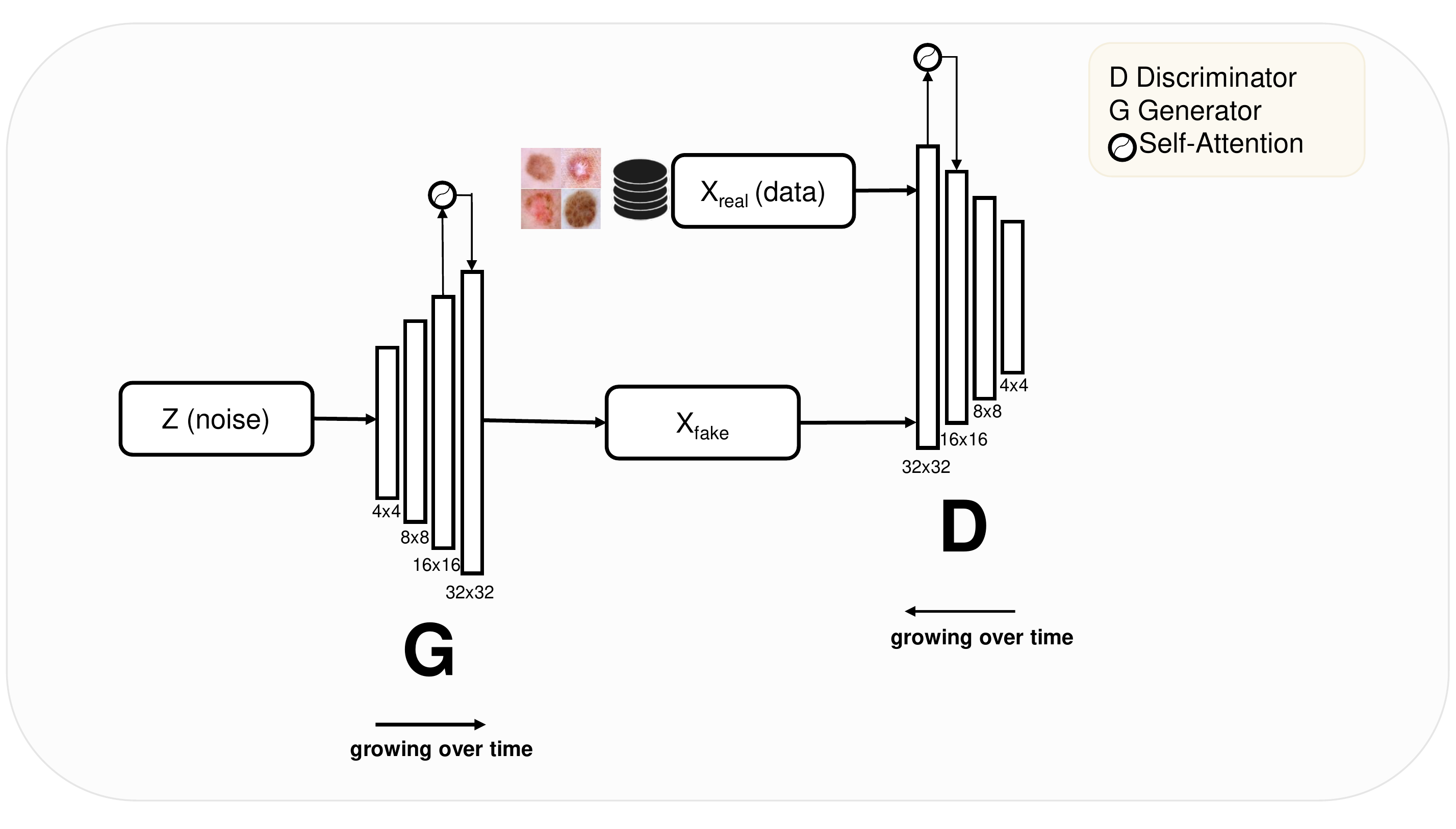}
	\caption{An overview of the APGAN+TTUR employed for skin lesion synthesis.}\label{fig:APGAN}	
\end{figure}

\begin{figure}[!ht]
	\centering
	\captionsetup{width=\columnwidth}
	\includegraphics[width=\columnwidth]{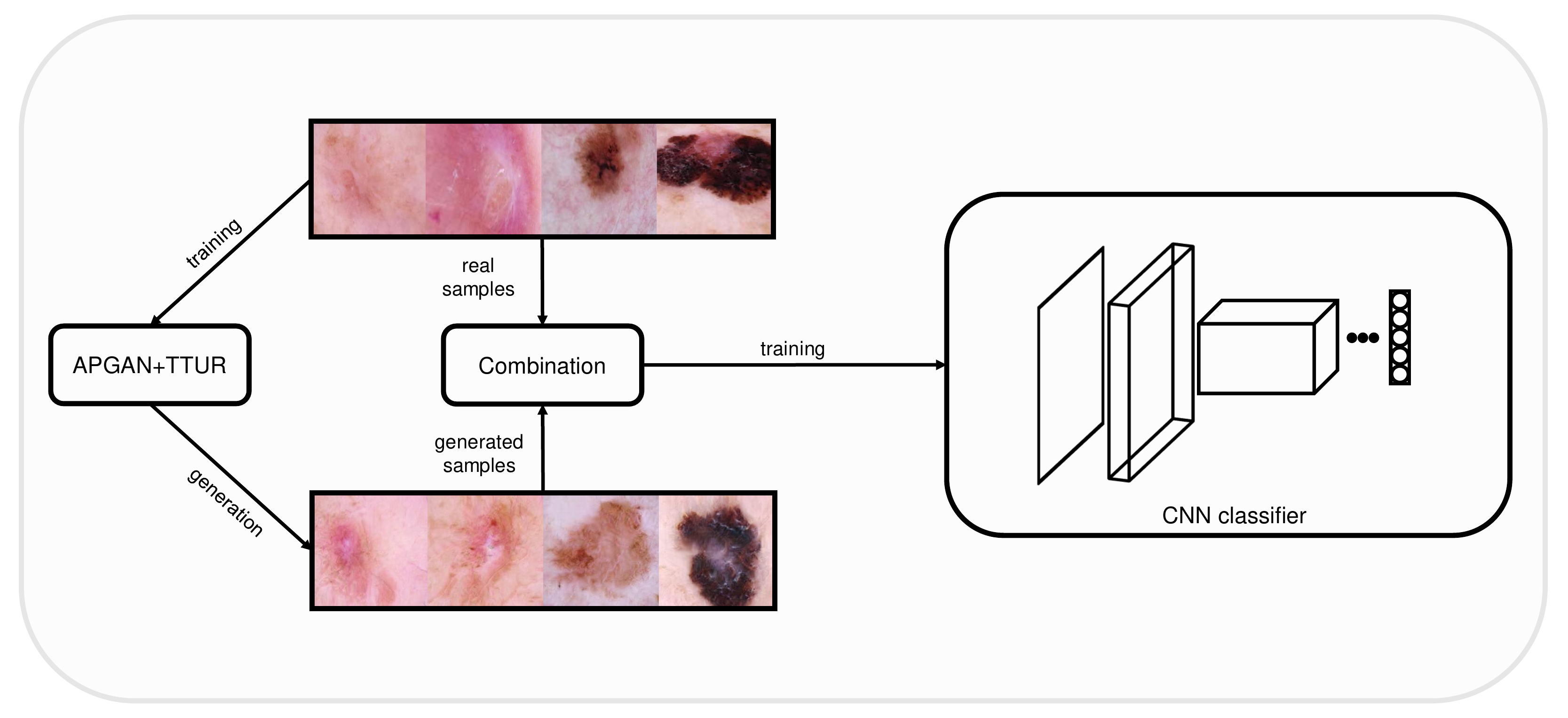}
	
	\caption{An overview of the overall process. The training set of ISIC 2018 dataset is used to train The APGAN+TTUR framework. Real samples and generated samples of APGAN+TTUR are combined to train the CNN classifier. }\label{fig:overall_framework}
\end{figure}

In order to tackle the imbalanced class problem, we have to automatically find class-preserving transformations for generating a valid and representative samples to boost further the classification accuracy. However, for skin cancer classification, the samples must have a higher level of detail (high resolution) in order to spot the presence of malignancy markers and their fine-grained details that differ a benign from a malignant skin lesion. To automatically find class-preserving transformations and generate skin lesion images of high resolution $256\times 256$ pixels, we propose APGAN+TTUR framework as shown in Figure \ref {fig:APGAN}. The APGAN+TTUR framework is based on following aspects (i) Progressive growing of GANs (ii) Self-Attention and (iii) Two Time Update Rule.  The overall process is illustrated in Figure \ref {fig:overall_framework}.

\subsection{Progressive Training of GANs}
The research towards using GANs has recently led to a breakthrough for synthesizing ever-increasing resolution of images in the work of \cite {karras2017progressive}. The underlying idea is to facilitate high-resolution image synthesis from noise at unprecedented levels of quality and realism. The output-resolution of the generator and the input-resolution to the discriminator are simultaneously ramped up by gradually adding new layers to the generator and discriminator networks leading to a very stable training behavior and very realistic, synthetic images at resolutions up to $1024 \times 1024$ pixels. Progressive training reduces training time, since most of the iterations are done at lower resolutions where the network sizes are small. The original work includes several further important contributions. A dynamic weight initialization method is proposed to equalize the learning rate between parameters at different depths, batch normalization is substituted with a variant of local response normalization in order to constrain signal magnitudes in the generator, and a new evaluation metric is proposed (Sliced Wasserstein distance). Our APGAN and APGAN+TTUR frameworks utilize the PGAN architecture, since it has shown outstanding results at generating images of high-resolution with a minimum number of parameters.

\subsection{Self-Attention Progressive GAN}
\label{sec:Self_Attention_Progressive_GAN_sec}
\begin{figure}[!ht]
	\centering
	\captionsetup{width=\columnwidth}
	\includegraphics[width=\columnwidth]{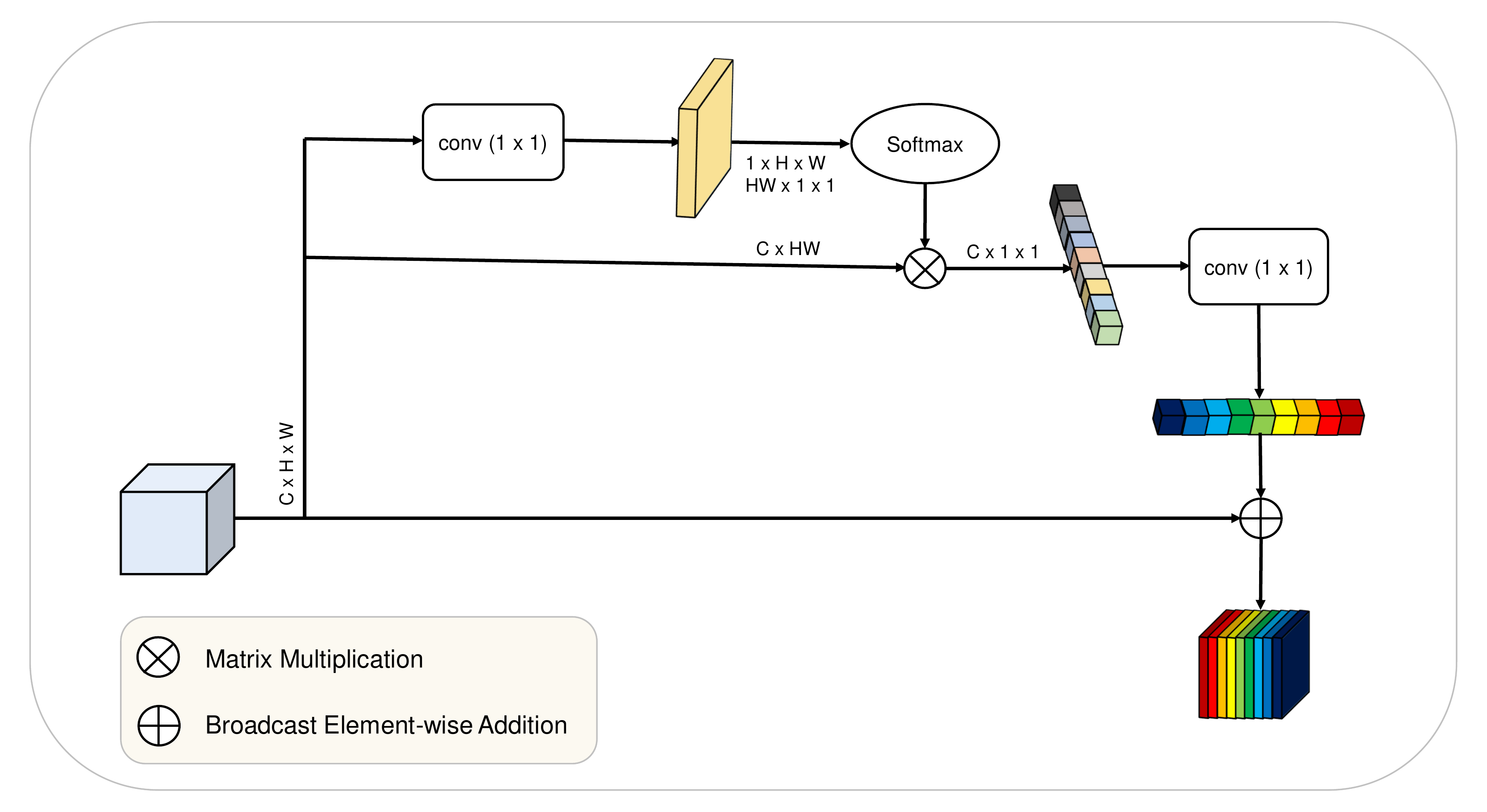}
	\caption{Simplified NL block (Equation \ref{eq_final_sim_non_block}).}\label{fig:SNL}
\end{figure}

Self-attention mechanism is a widely used mechanism in various tasks, such as machine translation \cite { gehring2016convolutional, gehring2017convolutional, vaswani2017attention}, graph embedding \cite{ velivckovic2017graph}, generative modeling \cite{ zhang2018self}, and visual recognition \cite{ wang2017residual, hu2018relation, wang2018non, yuan2018ocnet}. The basic building block of all the state-of-the-art architectures in computer vision consists of the convolution operation which is stacked in multiple layers to learn a hierarchy of features. These representations are learned over a series of convolution operations, however, due to the physical design of convolutional filters, the information flow in convolutional neural networks is restricted inside local neighborhood regions, which limit the overall understanding of complex scenes. This problem can be seen in \cite{radford2015unsupervised} where convolution layers are mainly used for image generation purpose.  All of these experiments have one thing in common the lack of convolution operations to capture geometrical shapes. For example, four-legged animals demands long range dependencies in the generator because of its complex contour. Recently, Goodfellow et al.\cite {zhang2018self} incorporated a self-attention mechanism which acts complimentary to convolution operation. Furthermore, Brock et al.\cite {brock2018large} used self-attention mechanism for high-fidelity natural image synthesis improving the state-of-the-art Inception score (IS) and Frechet Inception distance (FID) from 52.52 to 166.5 and 18.65 to 7.4.
\\ \indent
The non-local block can be deemed as a global context modeling block, which aggregates query-specific global context features (weighted averaged from all positions via a query-specific attention map) to each query position. As attention maps are computed for each query position, the time and space complexity of the non-local block are both quadratic to the number of positions $N_{p}$. Mathematically, the non-local block can be expressed as 
\begin{equation}
\label{eq_non_block}
z_{i} = x_{i} + W_{z}\sum_{j=1}^{N_{p}} \frac{f(x_{i}, x_{j})}{C(x)} (W_{v} \cdot x_{j}),
\end{equation}
where $i$ is the index of query positions, and $j$ enumerates all possible positions. $f(x_{i}, x_{j})$ denotes the relationship between position $i$ and $j$, and has a normalization factor $C(x)$. $W_{z}$ and $W_{v}$ denote linear transform matrices (e.g., 1x1 convolution). For simplification, we denote $ w_{ij} = \frac{f(x_{i}, x_{j})}{C(x)}$ as normalized pairwise relationship between position $i$ and $j$. The observation of \cite{cao2019gcnet} that the attention maps for different query positions are almost the same, they simplify the non-local block by computing a global (query-independent) attention map and sharing this global attention map for all query positions. They omit $W_{z}$ in the simplified version. Hence the simplified non-local block (SNL) is defined as
\begin{equation}
\label{eq_sim_non_block}
z_{i} = x_{i} + \sum_{j=1}^{N_{p}} \frac{exp(W_{k}x_{j})}{\sum_{m=1}^{N_{p}}exp(W_{k}x_{m})} (W_{v} \cdot x_{j}),
\end{equation}
where $W_{k}$ and $W_{v}$ denote linear transformation matrices. They also reduce the computational cost of Equation \ref{eq_sim_non_block} by applying the distributive law to move $W_{v}$ outside of the attention pooling, as
\begin{equation}
\label{eq_final_sim_non_block}
z_{i} = x_{i} + W_{v}\sum_{j=1}^{N_{p}} \frac{exp(W_{k}x_{j})}{\sum_{m=1}^{N_{p}}exp(W_{k}x_{m})}x_{j}.
\end{equation}
Equation \ref{eq_final_sim_non_block} is illustrated in Figure \ref{fig:SNL}.
\\ \indent
Our approach leverages the block of Equation \ref{eq_final_sim_non_block} to introduce self-attention to the PGAN architecture. Incorporating self-attention mechanism teaches the PGAN to focus on target structures of varying shapes and sizes, in other words, the discriminator implicitly learns to suppress irrelevant regions in an input image while highlighting salient features useful for a specific task which leads the generator to generate images with fine-grained and high-quality images. An overview of the setup is given in Figure \ref{fig:APGAN}. The self-attention is incorporated before the Downsample layer of discriminator and after the Upsample layer of generator.

\subsection{Two Time Update Rule}
\label{Two_Time_Update_Rule}
Despite using progressive training, we still had to overcome notable stability issues, due to the high resolution. Two Time Update Rule (TTUR) is used which is a stabilization technique for GAN training which improves both quantitative and qualitative results as proved in \cite{zhang2018self}, thus, we set the learning rate for the discriminator 5x compared to the generator while keeping the discriminator to generator step ratio as 1:1.


\section{Experiments and Results}
\label{sec:Experiment}

In the first part of our experiments, we examine the proposed self-attention mechanism in Section \ref{sec:Self-attention_mechanism_subsec}. Next, the effectiveness of the TTUR for stabilizing GAN training is evaluated in Section \ref{sec:High_Resolution_Skin_Lesions}. In the second part of our experiments, we examine the utility of GANs for data augmentation, i.e., for generating additional training samples, with the best-performing GAN model to boost classification accuracy.

\subsection{Dataset}
We evaluate our method on the ISIC2018 dataset that consists of 10015 skin lesion images from seven skin diseases- Melanoma (1113), Melanocytic nevus (6705), Basal cell carcinoma (514), Actinic keratosis (327), Benign keratosis (1099), Dermatofibroma (115) and Vascular (142). The megapixel dermoscopic images are center cropped to $450\times450$ pixels and then downsampled to $256\times256$ pixels. Figure \ref{fig:APGAN_gen_imgs} part (a) shows some of these training samples. We split the data into train (9514 images) and validation (501 images).
\\ \indent
For training PGAN, APGAN and APGAN+TTUR, We augment the training set to boost the GAN performance. We utilize rotation (in the range of [$-90\degree$, $90\degree$]), horizontal and vertical flipping, scale and skew. A python package named Augmentor \cite{bloice2017augmentor} was used for the augmentation.
\\ \indent
For training the classification network, we use the training set (9514 images) and the generated samples of GAN. For validation, the validation set is used (501 images). The detailed training process is investigated in Section \ref{sec:GAN_data_augmentation}.  

\subsection{Evaluation Metrics}
\begin{figure}[!ht]
	\fcapside
	{\caption{Illustration of GAN-train and GAN-test. GAN-train learns a classifier on GAN generated images and measures the performance on real test images. So, GAN-train evaluates the diversity and realism of GAN images. GAN-test learns a classifier on real images and evaluates it on GAN images. This measures how realistic GAN images are.} \label{fig:GAN_measures_fig}}
	{\includegraphics[keepaspectratio=TRUE,width=\columnwidth]{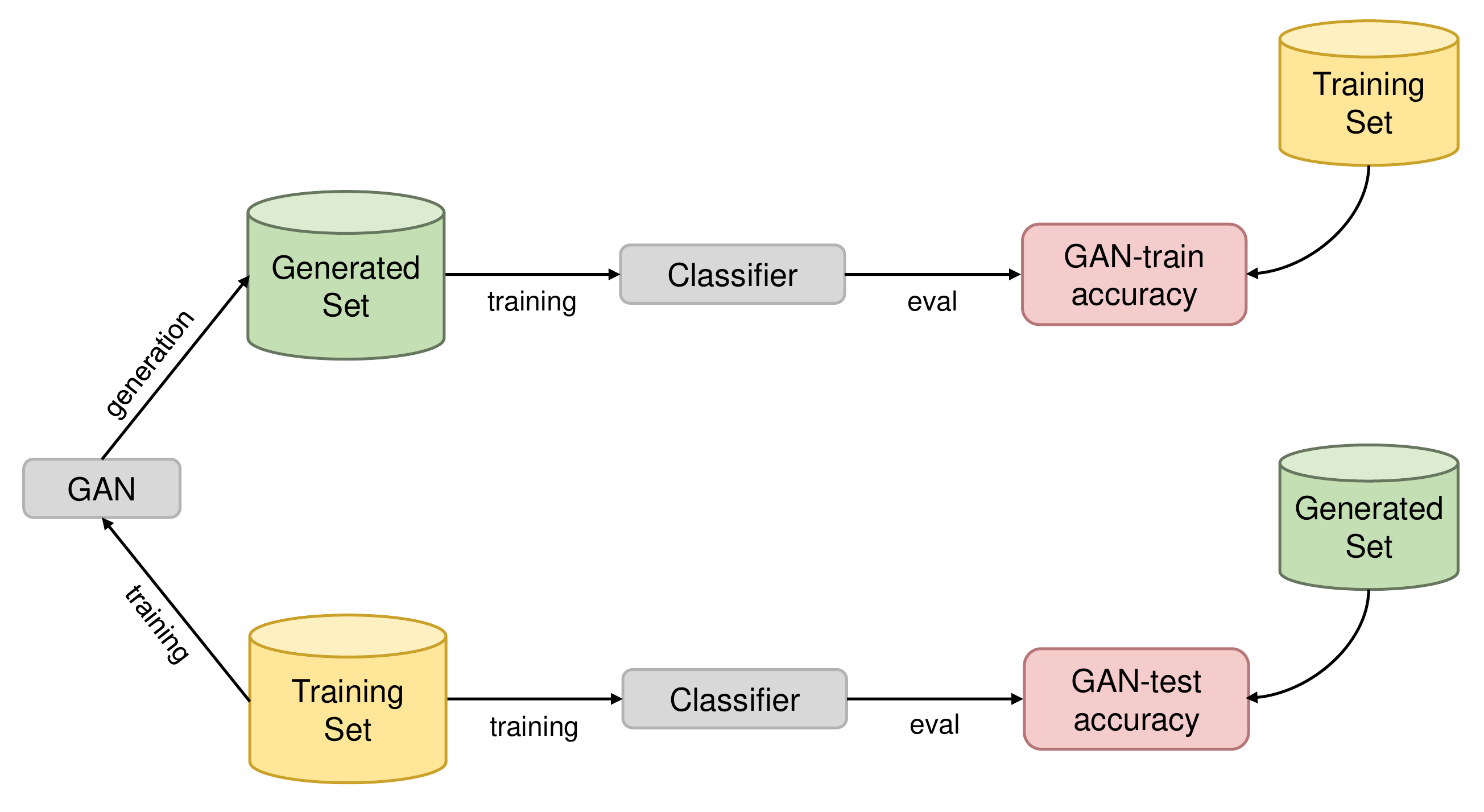}}
	
\end{figure}

A variety of methods have been proposed for evaluating the performance of GANs in capturing data distributions and for judging the quality of synthesized images. In order to evaluate visual fidelity, numerous works utilized either crowdsourcing or expert ratings to distinguish between real and synthetic samples. There have also been efforts to develop quantitative measures to rate realism and diversity of synthetic images. The two Inception-based, the Inception score (IS) \cite{salimans2016improved} and the Fréchet Inception distance (FID) \cite{heusel2017gans}, are useful measures to evaluate how training advances, but they guarantee no correlation with performance on real-world tasks they are also insufficient to finely separate state-of-the-art GAN. In addition, we noticed that they do not provide meaningful scores for skin lesions as the GoogleNet focuses on the properties of real objects and natural images. Shmelkov et al. \cite{shmelkov2018good} proposed two measures based on image classification, GAN-train and GAN-test, to compare class-conditional GANs. GAN-train and GAN-test measure the quantitative (diversity) and qualitative (quality of the image) of GANs respectively. GAN-train is the accuracy of a network trained on GAN generated images and is evaluated on real-world images, whereas GAN-test is the accuracy of a network trained on real images and evaluated on the generated images.
\\ \indent
Intuitively, when GAN-train accuracy is close to validation accuracy, it means that GAN images are as diverse as the training set. On the other hand, when GAN-test accuracy is close to validation accuracy, it means that GAN does capture the target distribution well and the image quality is good. For illustration, see Figure \ref {fig:GAN_measures_fig}.

\subsection{Experimental Setup}
\label{sec:Experimental_Setup}
All of our experiments are performed on NVIDIA 1080Ti with 24GB main memory. For GAN experiments, the conditional version of PGAN (official TensorFlow implementation) is modified using state-of-the-art tweaks like (i) Self-Attention and (ii) 5:1 learning rate ratio between discriminator and generator (TTUR). To determine the best attention placement, several experiments are conducted at resolution $128 \times 128$ pixels as discussed in Section \ref{sec:Self_Attention_Progressive_GAN_sec}. To generate skin lesions of high resolution $256 \times 256$ pixels, TTUR is utilized alongside self-attention mechanism. APGAN is trained for $13.5$ hours, whereas APGAN+TTUR is trained for $31.7$ hours.

The classifier for evaluating GAN-train and GAN-test is ResNet-18 \cite{he2016deep} using PyTorch framework. It initialized with weights pretrained on Imagenet. The model is trained for 50 epochs using a momentum optimizer with learning rate 0.001 (using a batch size of 64). However, Resnet-18 takes $256 \times 256$ images as input. To utilize ResNet-18 with images of size $128 \times 128$, AvgPool2d is replaced by AdaptiveAvgPool2d. The images are loaded in to a range of [0, 1] and then normalized using mean = [0.485, 0.456, 0.406] and std = [0.229, 0.224, 0.225].

\subsection{Self-Attention mechanism}
\label{sec:Self-attention_mechanism_subsec}
To inspect the effect of enhancing PGAN with self-attention mechanism, we build several APGAN models, at $128\times128$ pixels, by incorporating the self-attention mechanism to different stages of the generator and discriminator. As shown in Table \ref{tab:stages_ex_table}, the APGAN models with the self-attention mechanism at the $2^ {N-1} $-to-$2^ {N} $ level feature maps (e.g., stage 64 and stage 128) achieve better performance than the models with the self-attention mechanism at the low level feature maps (e.g., stage 32 and stage 64). For example, the GAN-train of the model PGAN is improved from 67.7 to 70.1 by “APGAN, feat 64”. Moreover, GAN-test of the model PGAN is improved from 60.8 to 62.2 by “APGAN, feat 64”.  The reason could be that the attention coefficients result from large feature maps learn to highlight salient image regions that are passed through $2^{N-1}$-to-$2^{N}$ level feature maps and prune low-level feature responses to maintain only the activations relevant to the specific task. The attention mechanism gives more power to both generator and discriminator leading the PGAN to generate better quantitative and qualitative synthetic samples In addition, the comparison of our APGAN and the PGAN (3rd column of Table \ref{tab:stages_ex_table}) demonstrate the effectiveness of enhancing the PGAN with self-attention mechanism.

\begin{table}[!ht]
	\centering
	\captionsetup{ margin=1.5cm}
	
	\begin{center}
		\begin{tabular}{|c|c|c|c|c|c|}
			\hline
			\multirow{2}{1.5cm}{\centering Model}&
			\multirow{2}{1.5cm}{\centering Real Images}&
			\multirow{2}{1.5cm}{\centering PGAN}& \multicolumn{3}{c|}{ APGAN} \\ 
			\cline{4-6}& & &  \multicolumn{1}{c|}{Stage 32} & \multicolumn{1}{c|}{Stage 64} & \multicolumn{1}{c|}{Stage 128} \\ \hline
			GAN-train & 78.2& 67.7 & 63.3 & \textbf{70.1} & 64.7 \\ \hline
			GAN-test & - &60.8 & 56.0 & \textbf{62.2} & 58.1 \\ \hline
		\end{tabular}
	\end{center}
	\caption{ Comparison of APGAN and PGAN. The SNL block is added to different stages of the generator and discriminator. All models have been presented with over 6M images, and the best GAN-train and GAN-test are reported.}\label{tab:stages_ex_table}
\end{table}


\subsection{High Resolution Skin Lesions}
\label{sec:High_Resolution_Skin_Lesions}
For better classifying the presence or absence of malignancy, the skin lesion is synthesized at $256\times256$ pixels. In order to generate $256\times256$ of skin lesions, the PGAN, APGAN and APGAN+TTUR have been presented with 8 million images, which is equivalent to over 3.2M iterations. We use a minibatch size 256 for resolutions $4^ {2}–128^ {2} $ and then gradually decrease the size according to $16 \rightarrow 128$, $32 \rightarrow 64$, $64 \rightarrow 32$, $128 \rightarrow 16$, and $256 \rightarrow 8$. Despite using progressive training and the incorporating of self-attention mechanism, the generated samples suffer from some artifacts, due to the high resolution as discussed in Section \ref{sec:Artifacts}. To mitigate unstable training behavior, the imbalanced learning rates (TTUR) is utilized. We set the learning rate for the discriminator is 0.004 and the learning rate for the generator is 0.001. To finely separate the best performance GAN, we use GAN-train and GAN-test of \cite {shmelkov2018good}. Training was resumed for an additional 3 network checkpoints. Per model, we generate 1000 synthetic images for each class and we measure GAN-train and GAN-test for each model. Results from each of the best models are reported. TTUR greatly stabilize APGAN training and improves both quantitative and qualitative results as shown in Table \ref {tab:ISIC_res_table}. Clearly, the best quantitative and qualitative results are obtained with the APGAN+TTUR samples.

\begin{table}[!ht]
	\centering
	\captionsetup{justification=justified, margin=3cm}
	\begin{tabular}{|c|c|c|}
		\hline
		Model       & GAN-train       & GAN-test \\ \hline
		Reals       & 84.6            & -        \\ \hline
		PGAN        & 62.5            & 56.2        \\ \hline
		APGAN       & 66.3            & 55.6        \\ \hline
		APGAN+TTUR  & \textbf{69.5}   & \textbf{60.0}        \\ \hline
	\end{tabular}
	\caption{Results from each of the best models are reported. We finetune the whole model of ResNet-18 to measure GAN-train and GAN-test.}\label{tab:ISIC_res_table}
\end{table}


\subsection{Artifacts}
\label{sec:Artifacts}

\begin{figure}[!ht]
	\centering
	\captionsetup{width=\columnwidth}
	\includegraphics[width=0.7\columnwidth]{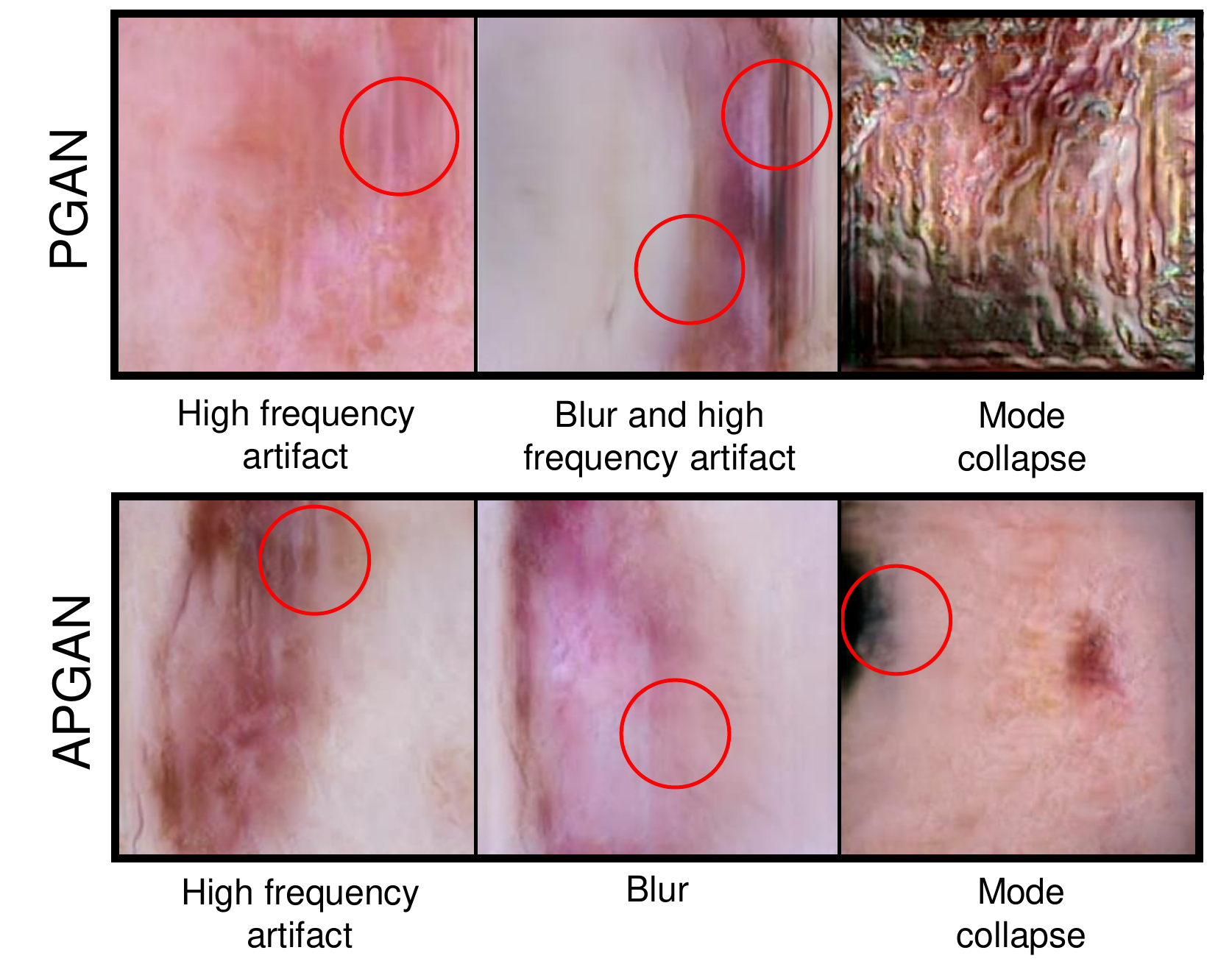}
	\caption{Most commonly seen artifact patterns in the generated samples of PGAN and APGAN. }\label{fig:PGAN_APGAN_Artifacts}
\end{figure}

\begin{figure}[!ht]
	\begin{center}
		\begin{tabular}{c}
			\includegraphics[height=5.5cm]{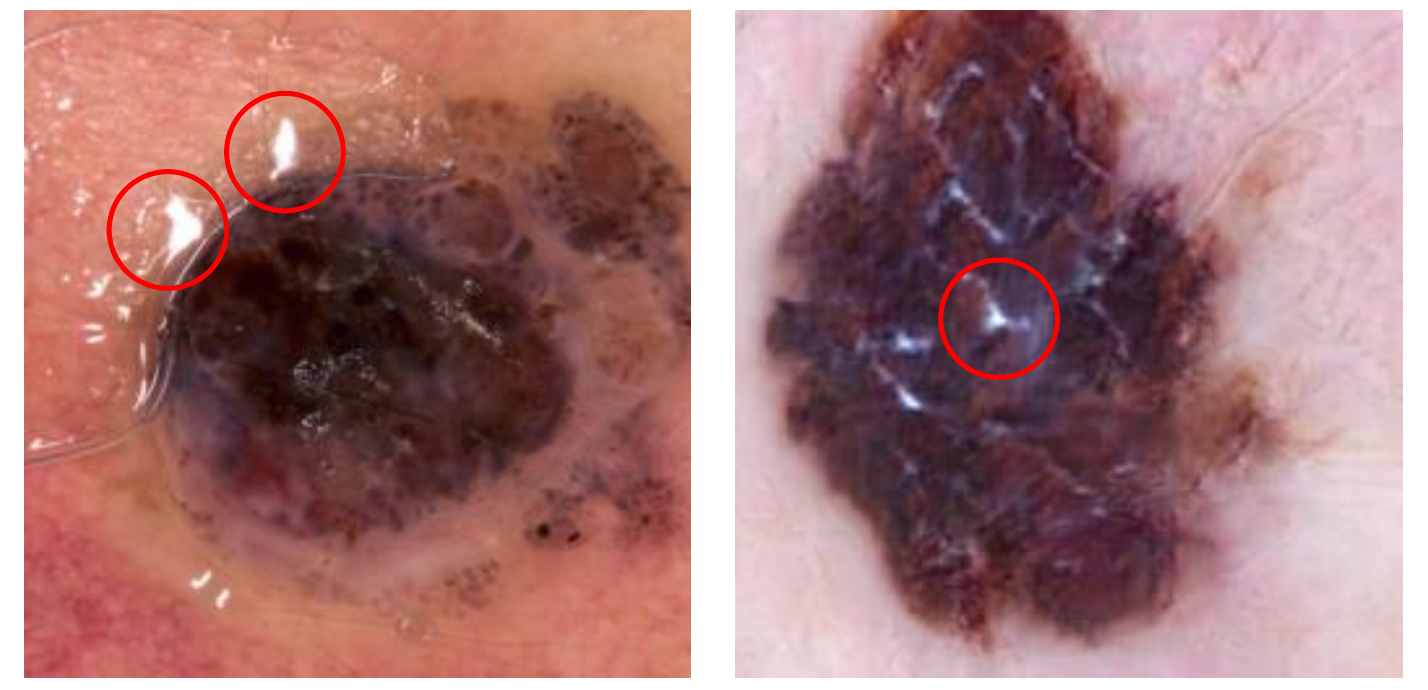}  
			\\
			(a) \hspace{4.8cm} (b)
		\end{tabular}
	\end{center}
	\caption 
	{ \label{fig:real_APGAN_TTUR_Artifacts}
	Note that bright spot artifact appear in figure (b) . This problem can be attributed to problem in the training set as shown in figure (a).} 
\end{figure}

Due to unstable training behavior, several types of artifacts are observed in the generated samples. For example, blur, high frequency artifacts and even mode collapse. As shown in Figure \ref{fig:PGAN_APGAN_Artifacts}. The unstable training behavior was mitigated by applying TTUR. However, some of APGAN+TTUR samples suffer from bright spot in melanoma class. This problem can be attributed to problem in the training set, since real samples of melanoma class have the same bright spot. Examples of such images are shown in Figure \ref{fig:real_APGAN_TTUR_Artifacts}.

\subsection{GAN data augmentation}
\label{sec:GAN_data_augmentation}
\begin{figure}[!ht]
	\centering
	\captionsetup{width=\columnwidth}
	\includegraphics[width=\columnwidth]{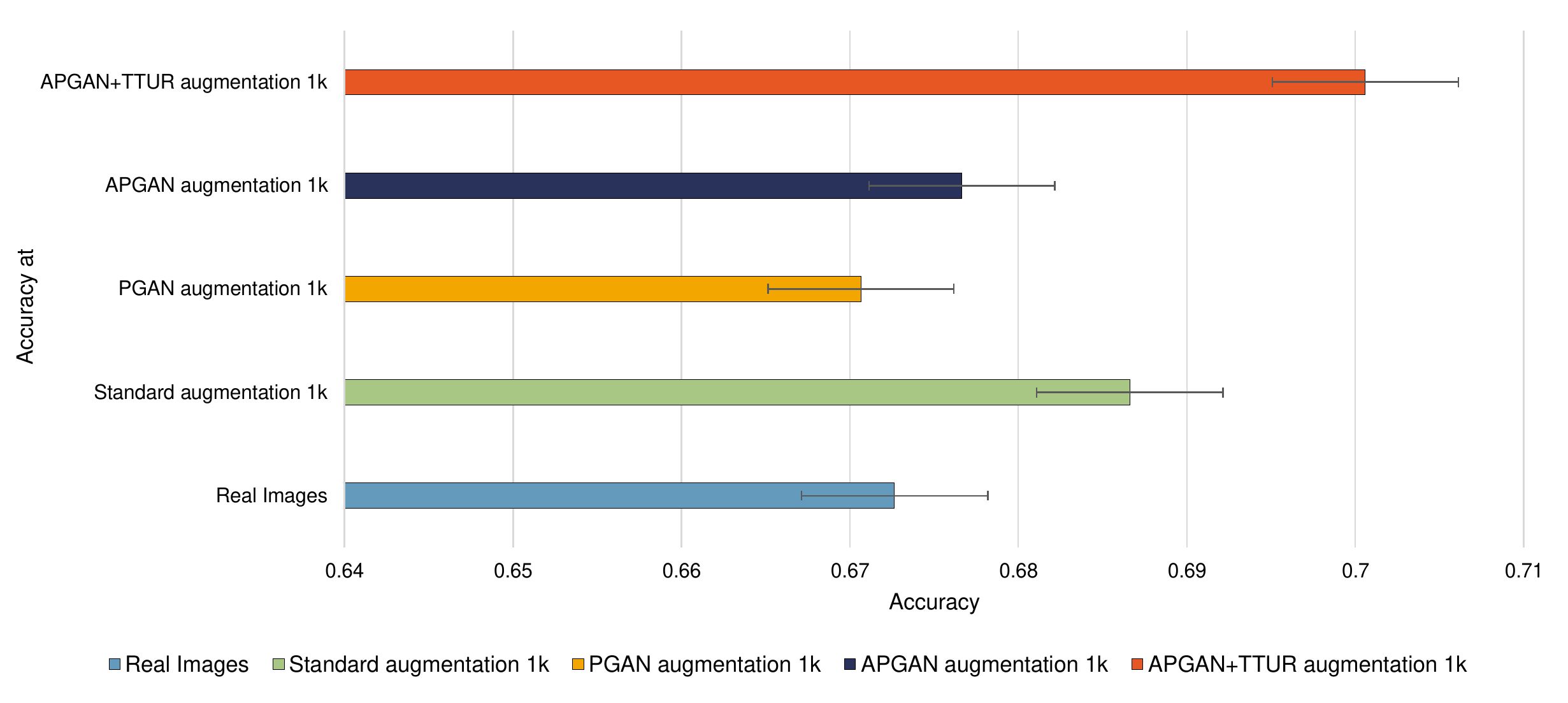}
	\caption{Multi-classification performance for training set with 100 images for each class and for augmenting each class with 1k images using APGAN+TTUR and standard augmentation methods. ResNet-18 is used as a classifier on validation set (501). These results are reported after 50 epochs.}\label{fig:Accuracy_fig}	
\end{figure}

In this section the training set (9514) is augmented using the best performed GAN (APGAN+TTUR) and standard augmentation methods. We randomly pick 100 images from each class of training set. The 100 images of each class are increased with 1k images using APGAN+TTUR and standard augmentation methods. For classification, ResNet-18 \cite{he2016deep} is utilized. It is configured as discussed in Section \ref{sec:Experimental_Setup}. The results are shown in Figure \ref {fig:Accuracy_fig}. We observe that the model trained on real images (100 images per class) achieves 67.3\% on validation set (501), while augmenting real images using APGAN+TTUR and standard augmentation methods achieved 70.1\% and 68.7\% respectively. Consequently, augmentation using APGAN+TTUR improves the accuracy over the corresponding real-only and standard augmentation counterparts by 2.8\% and 1.4\% respectively. Based on These results, two points can be concluded (i) the generated samples have clinically-meaningful features, since there is an information gain in the synthetic samples which improves the classification accuracy and (ii) standard augmentation methods use a limited set of known invariances, whereas APGAN+TTUR automatically learns a much broader invariance space. To prove further the superiority of APGAN+TTUR, real images were augmented with 1K samples using PGAN and APGAN. The results are 67.1\% and 67.7\% respectively.

\section{Conclusion.}
\label{sec:Conclusion}
In this paper we have proposed a novel enhancement using self-attention based progressive GAN framework for generating high-definition, visually-appealing and clinically-meaningful synthetic skin lesion images. The proposed model leverages state-of-the-art tweaks like (i) progressive growing of GANs (ii) Self-Attention and (iii) imbalanced learning rate (TTUR). Self-attention guides the discriminator to pay more attention to the presence of malignancy which results in making the generator to generate samples that contain fine-grained details to fool the discriminator. Despite the using of self-attention mechanism, the generated samples suffer from some artifacts due to unstable training behavior. The imbalanced learning rate (TTUR) is used to tackle this issue. Finally, APGAN+TTUR was utilized to generate additional training samples to boost further the classification accuracy. Noteworthy, there is an information gain in the synthetic samples, and consequently the classification accuracy is improved. Moreover, data augmentation using APGAN+TTUR has higher information gain than using standard data augmentation methods. In future work, we aim to utilize APGAN+TTUR for performing large scale experiments on multiple datasets.

%
%



\bibliographystyle{elsarticle-num}

\bibliography{sample}

\end{document}